\documentclass[final,authoryear,1p,times]{elsarticle}

\usepackage{amssymb}
\usepackage[latin1]{inputenc}
\usepackage{hyperref}
\usepackage{amstext}

\usepackage{color}
\definecolor{greencolor}{rgb}{0,0.5,0.2}
\definecolor{redcolor}{rgb}{.7,0.,0.}
\definecolor{bluecolor}{rgb}{0,0.,1.}
\definecolor{greycolor}{rgb}{.5,.5,.5}

\journal{Journal of Informetrics}

\begin{document}

\begin{frontmatter}



\title{Three-feature model to reproduce the topology of citation networks and the effects from authors' visibility on their h-index}


\author{Diego Raphael Amancio\footnote{Corresponding author: diego.amancio@usp.br, diegoraphael@gmail.com}, Osvaldo Novais Oliveira Jr., Luciano da Fontoura Costa}
\address{Institute of Physics of S\~ao Carlos\\
University of S\~ao Paulo, P. O. Box 369, Postal Code 13560-970 \\
S\~ao Carlos, S\~ao Paulo, Brazil \\}

\begin{abstract}
Various factors are believed to govern the selection of references in citation networks, but a precise, quantitative determination of their importance has remained elusive. In this paper, we show that three factors can account for the referencing pattern of citation networks for two topics, namely ``graphenes'' and ``complex networks'', thus allowing one to reproduce the topological features of the networks built with papers being the nodes and the edges established by citations. The most relevant factor was content similarity, while the other two - in-degree (i.e. citation counts) and {age of publication} had varying importance depending on the topic studied. This dependence indicates that additional factors could play a role. Indeed, by intuition one should expect the reputation (or visibility) of authors and/or institutions to affect the referencing pattern, and this is only indirectly considered via the in-degree that should correlate with such reputation. Because information on reputation is not readily available, we simulated its effect on artificial citation networks considering two communities with distinct fitness (visibility) parameters. One community was assumed to have twice the fitness value of the other, which amounts to a double probability for a paper being cited. While the h-index for authors in the community with larger fitness evolved with time with slightly higher values than for the control network (no fitness considered), a drastic effect was noted for the community with smaller fitness.
\end{abstract}


\end{frontmatter}


\section{Introduction}

Quantitative evaluations of researchers, institutions, geographical regions, journals and areas of science and technology have become commonplace, especially with the widespread availability of information in scientific databases. Citation counts and impact factors are among the most common parameters used and may be key for deciding on promotions, grants and identification of scientific trends. Science has become to a certain extent driven by scientometry~\citep{ref2,whatfactors,ref1}, which is motivation for detailed studies of the way scientometric parameters are defined and of patterns of citations~\citep{patterns}. Citation networks, for instance, have been modeled with concepts and methodologies of complex networks~\citep{ref14,newmanbook,ref15,ref17,ref16}. {The degree of these networks (i.e. the number of citations received by papers) was found to follow the scale free behavior~\citep{scientificamerican,price}}, which amounts to say that the probability of a paper being cited was dependent on its citation counts~\citep{newmanbook,price}. Also known is that content similarity plays a role on the choice of references~\citep{evdoc}, even though the correlation with the most related papers has been found to be low~\citep{goodpractices}. Other factors considered to affect the citation pattern are the {age of publication}, since recent papers are more likely to be cited than old ones~\citep{aging2,aging}, the reputation of authors, journals and institutions, and even the authors' language as they affect the readability of papers~\citep{whatfactors}.

With the variety of possible factors, modeling citation networks has not been straightforward. Traditional models considering one feature at a time may be successful in explaining the dynamics of this feature, but could on the other hand miss out in important points on overlooked features~\citep{evdoc}. The preferential attachment model~\citep{ref14}, for instance, predicts the degree distribution of the networks, but fails to match the actual content similarity of real databases~\citep{evdoc}. Other methods also explain the degree distributions~\citep{evdoc} or clustering coefficient~\citep{cc}, but not the content similarity and distribution of the time difference between papers and their references. According to Ref.~\citep{evdoc} these features follow well-known distributions. The content similarity obeys a Gaussian-like distribution, while the {age} dependence distribution follows a power law~\citep{pwlaw}. Therefore, in the attempts to model citation networks one should consider as many features as possible. In this paper, we propose a model that takes into account three factors believed to affect the pattern of citations, namely the in-degree distribution, the content similarity and the {age of publication}. We shall show that this model is capable of reproducing topological characteristics of citation networks obtained for two topics in the arXiv\footnote{http://www.arXiv.org} repository. Because it is hard to quantify the reputation or visibility of journals or institutions, this factor could not be included in the model. Alternatively, we designed artificial networks with two communities of authors differing in their visibility (fitness), i.e. with different probabilities of having their papers being cited. We shall show that differences in fitness cause major effects on the temporal evolution of h-index~\citep{ref2,hindex2,ref9} of authors.

\section{Modeling Citation Networks}

We propose a model to describe features of citation networks in which three parameters are assumed to govern the network, namely topology, content similarity (semantics) and {age of publication}. Simulated networks were then created with the citations being selected according to one of these criteria, and then with a combination of the three criteria. The content similarity was computed by collecting papers from the arXiv repository for two topics, viz. ``complex networks'' and ``graphene'', yielding the networks referred to as CN and GF, respectively. For the sake of processing times, only the abstracts were considered, and each paper was characterized by the frequency of lemmatized\footnote{The lemmatization consists in converting words to their canonical form. In this step, verbs are converted into their infinitive form and nouns are converted into their singular form.} words, disregarding stopwords\footnote{Stopwords are highly frequent words conveying little semantic meaning, such as articles and prepositions.}. Assuming that the frequency of words in papers $a$ and $b$ are given by the vectors $\overrightarrow{v_a}$ and $\overrightarrow{v_b}$, where the element $\overrightarrow{v}(i)$ represents the frequency of word $i$, then the content similarity $\sigma_{ab}$ between the two papers is:
\begin{equation}
\sigma_{ab} = \frac{\overrightarrow{v_a}  \cdot \overrightarrow{v_b}}{\|\overrightarrow{v_a}\| \cdot \|\overrightarrow{v_b}\|}.
\end{equation}
Because $\sigma_{ab}$ gives the cosine of the angle between the vectors, $\sigma_{ab}$ lies between $0$ and $1$. As reported in Ref.~\citep{evdoc} and verified in both real networks extracted from arXiv, the distribution of $\sigma_{ab}$ for every $a$ citing $b$ follows a normal distribution:
\begin{equation}
p(\sigma_{ab}) = \frac{1}{\sqrt{2 \pi s^2}} \exp \Bigg{(} -\frac{(\sigma_{ab}-\mu)^2}{2s^2} \Bigg{)},
\end{equation}
where $\mu$ and $s^2$ are the mean and variance, respectively.

The other criterion to select the citations is a preferential attachment rule based on the current in-degree of a paper. Thus, papers with high citation counts are more likely to be cited again, according to a power law $p(k) \propto k^{-\gamma_k}$, where $k$ is the in-degree and $\gamma_k$ is the coefficient of the power-law $p(k)$, computed according to the methodology devised in Ref.~\citep{bauke}. As for the criterion of {age of publication}, the citation count is taken as inversely proportional to the time difference $\Delta t$ between the article and its references. As observed in Ref.~\citep{aging}, and confirmed in our $2$ real citation networks, the power law function $p(\Delta t) \propto \Delta t^{-\gamma_t}$ can be used to characterize the likelihood of an article being cited $\Delta t$ months after its publication date.

\begin{figure}[h]
\begin{center}
\includegraphics[width=0.5\textwidth]{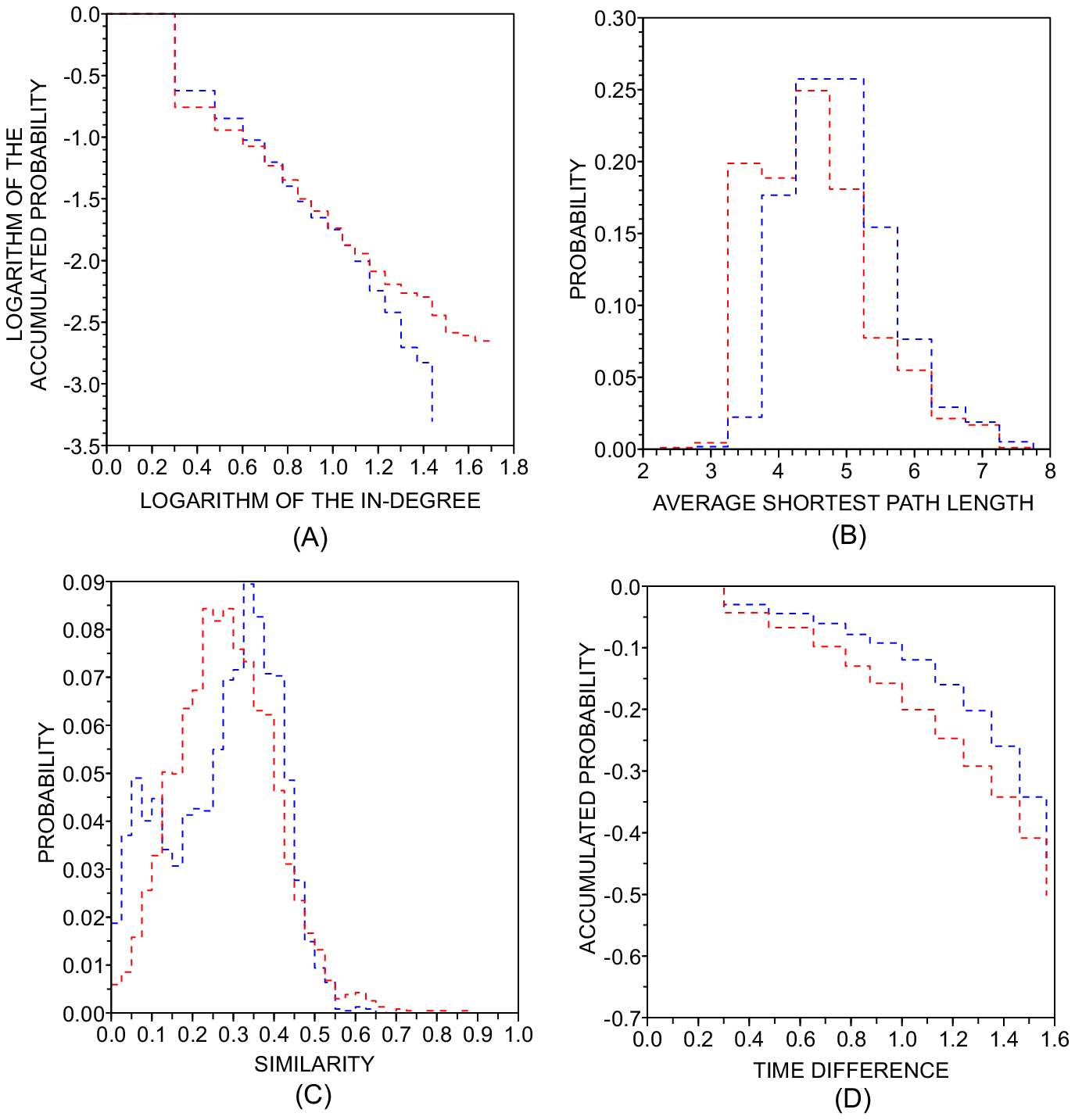}
\end{center}
\caption{\label{fig3} Distributions for the CN (real) network (red) and the proposed model (blue). The properties predicted by the model include topological (in-degree in graphic (a) and average shortest path length~\citep{ref17} in (b)), semantic (content similarity between a paper and its references) in (c)) and temporal (figure d)) features.}
\end{figure}

\begin{figure}[h]
\begin{center}
\includegraphics[width=0.5\textwidth]{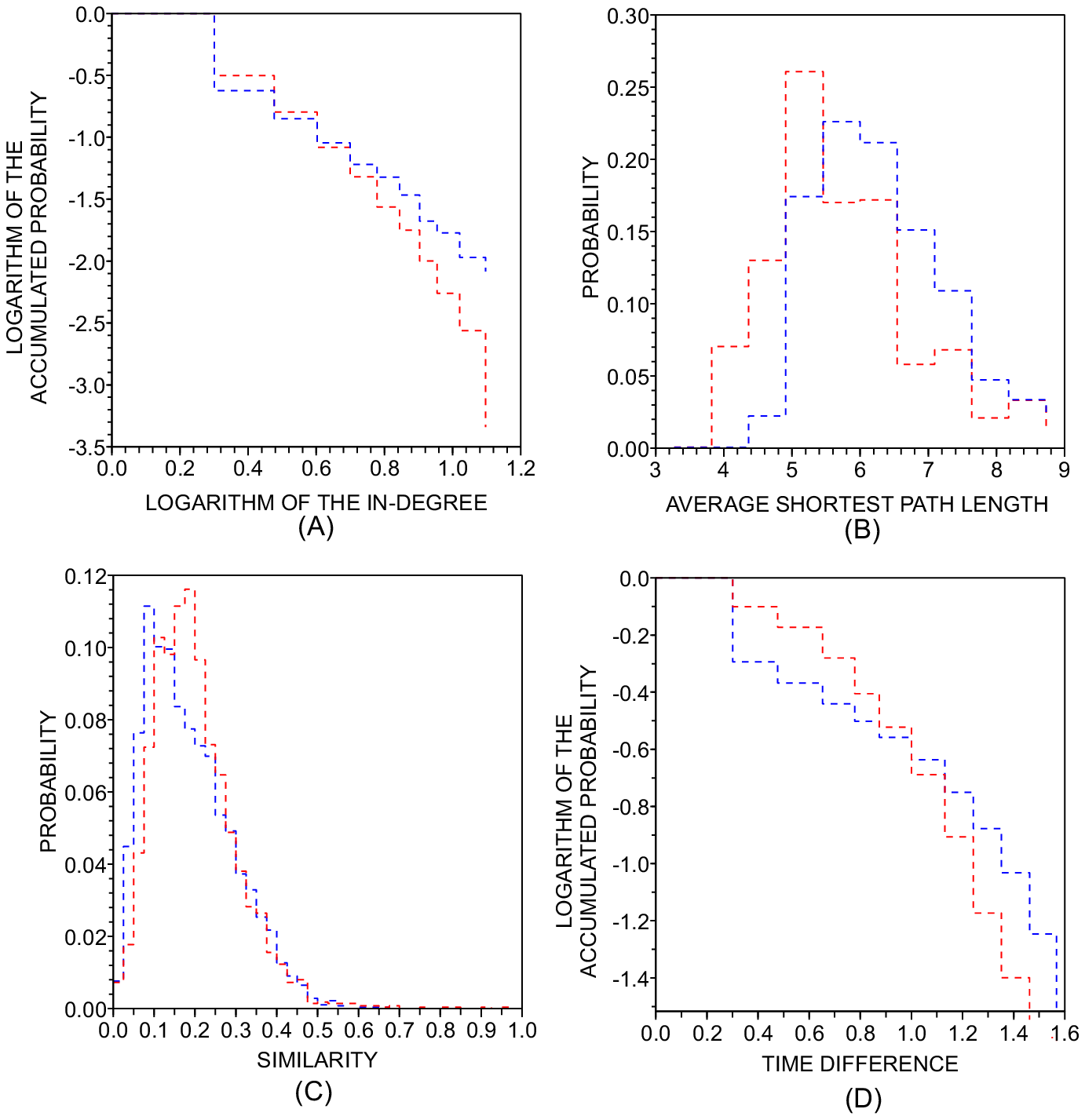}
\end{center}
\caption{\label{fig4} Distributions for the GF (real) network (red) and the proposed model (blue). The properties predicted by the model are the same as in Figure 1.}
\end{figure}

The simulated networks obtained with only one of the criteria exhibited topological properties that differed considerably from the real networks extracted from the arXiv repository for both subjects ``complex networks'' (CN) and ``graphene'' (GF) (results not shown). This finding is depicted quantitatively by determining the error $\epsilon^2$ (see definition in Appendix A) in Tables \ref{tab.2} and \ref{tab.3} in the attempt to fit the networks. Excellent agreement was observed, however, when the three criteria were combined in an optimization procedure, as shown in Figures \ref{fig3} and \ref{fig4} for the CN and GF networks. The contribution from each of the criteria ($\alpha$ for topology, $\beta$ for content similarity and $\lambda$ for time difference) was computed upon minimizing $\epsilon^2$, as described in Appendix A.

\begin{table}[h]
\centering
\caption{\label{tab.2} Best model found with the simulated annealing heuristic (see Appendix A). The combination of the {three criteria} gives optimized results, because in the best cases $\alpha$, $\beta$ and $\lambda \neq 0$. In other words, the model yielding the minimum error $\epsilon_{min}^2$ employs all the {three features}.}
\begin{tabular}{|c|c|c|c|c|}
\hline
\textbf{Network} & $\alpha$ & $\beta$ & $\lambda$ & $\epsilon_{min}^2$ \\
\hline       Complex Network  & 40.0 \% & 52.5 \% & 7.5 \% & 0.056\\
Graphene& 5.0 \% & 45.0 \% & 50.0 \% & 0.128\\
\hline
\end{tabular}
\end{table}

\begin{table}[h]
\centering
\caption{\label{tab.3} Models obtained with only one factor at a time and comparison with the minimum error $\epsilon_{min}^2$ obtained with the models depicted in Table \ref{tab.2}. Because $\epsilon^2 /  \epsilon_{min}^2 > 1$, the model combining the 3 factors is more accurate than those considering only one feature.}
\begin{tabular}{|c|c|c|c|c|}
\hline
\textbf{Network} & $\alpha$ & $\beta$ & $\lambda$ & $\epsilon^2 /  \epsilon_{min}^2 $ \\
\hline
Complex Network  & 100.0 \% & 0.0 \%   & 0.0 \%   & 3.125 \\
Complex Network  & 0.0   \% & 100.0 \% & 0.0 \%   & 2.104 \\
Complex Network  & 0.0   \% & 0.0 \%   & 100.0 \% & 7.982 \\
\hline
Graphene& 100.0 \% & 0.0 \% & 0.0 \%   & 2.617 \\
Graphene& 0.0   \% & 100.0 \% & 0.0 \% & 1.945 \\
Graphene& 0.0   \% & 0.0 \% & 100.0 \% & 3.445 \\
\hline
\end{tabular}
\end{table}

The results in Table \ref{tab.2} indicate that for both networks the similarity of content is an important criterion for selecting references, being responsible for approximately 50~\% of the citations. The preferential attachment (represented by taking the in-degree into account) was relevant for the CN networks, while the {age of publication} was more relevant for the GN network. Even though the content similarity is the most important factor, this does not mean that authors are selecting for the list of references the most similar papers to the manuscript being produced. This can be observed both in the distribution of figures \ref{fig3}(c) and \ref{fig4}(c), which show that only a few cited articles are very similar. It is also consistent with the low correlation found between the actual list of references and the most similar papers in a database in another piece of work~\citep{goodpractices}.

It has to be admitted, nevertheless, that the need to employ distinct parameters for reproducing the real networks indicates that the {three-criterion} model is not universal. It cannot account for all features of citation networks. This limitation was indeed expected since intuitively one knows that other criteria are important for selecting references. Perhaps the most relevant is the reputation (or visibility) of authors and journals~\citep{newref1,newref2}, which is partially (but not entirely) implicit in the in-degree incorporated in our model. We did not include the visibility criterion in the model because this type of information is not readily available. For example, not all papers in the arXiv database have been published, so it is impossible to use the impact factor of the corresponding journals. Regarding the institutions, there is no well established index quantifying their notoriety or reputation. As for the authors, use could be made of the ISI.highlycited.com database\footnote{http://researchanalytics.thomsonreuters.com/highlycited}, but only a small number of authors are listed.

We have decided to consider visibility in its possible effects on citation networks, which is performed in the next section.

\section{Effects from Visibility on the Evolution of h-index} \label{visibilidade}

 In order to analyze how visibility interferes on the dynamics of the citation network, we study the evolution of the h-index~\citep{ref2,hindex2,ref9} of authors belonging to two artificial communities with distinct visibility, assuming that the one community is twice as visible as the other one. Four models were considered differing in terms of in-degree distribution and fitness. In all models, we assume that the number of articles published by an author each year follows a power law $p(y) = c~y^{\gamma}$, where $p(y)$ represents the probability distribution of $y$, and $\gamma$ and $c$ are real parameters. These values were determined by defining the endpoints ($y$,$p(y)$) of the distribution: (1,$m$) and ($s$,1). Consequently, we assume that $m$ authors publish one paper every year and only one author publishes $s$ papers per year. With these limits, $c = m$ and
 \begin{equation}
 \gamma = \frac{\log(m)}{\log(s)}.
 \end{equation}
 In our experiments we assume $m = 15$ and $s = 30$. Note that from $p(y) = c~y^\gamma$, it is necessary to sample some values of $y$. Without loss of generality, we chose the following values: $y =$ (1,2,3,5,10,15,30) and obtained $p(y) =$ (15,9,6,4,2,2,1). In other words, $15$ authors are assumed to publish one article every year, $9$ authors publish two articles per year and so on. Therefore, $N_a = \sum p(y) = 39$ authors and a total of $N_p = \sum y p(y) = 151$ papers were published per year. This distribution was assumed for each one of the communities (hereafter referred to as communities A and B) and thus a total of $302$ papers were published by $78$ authors.

 The citation network was represented with a digraph $\Gamma = ($V$,$E$)$ where the vertices $V$ are papers and edges $E$ are established with citations between papers. Because the model is increased by incorporation of new papers over a period of $25$ years, both $V$ and $E$ increased with time. In order to distinguish communities A and B with regard to their visibility (i.e., the likelihood to receive new citations), we arbitrarily assumed the visibility (fitness) $f_A$ of community A as being twice the fitness $f_B$ of community B. That is to say, articles in community A are twice as likely to be cited. The different values of visibility were adopted to simulate differences arising due to distinct impact factors of journals or authors' institutions, among others~\citep{whatfactors}. The growth of the citation networks was obtained for each year.

  The four models used are: (i) UNI: uniform, random selection of references; (ii) PREF: preferential selection of references depending on the fitness of the community; (iii) PREFC: preferential selection for papers with larger in-degree (i.e. highly cited papers); and (iv) DBPREF: preferential selection depending on the fitness and in-degree. In the UNI model, each article included is assumed to cite $w$ randomly selected published papers. Analogously, in the PREF model random papers are cited, but considering the community visibility. The PREFC model is also preferential, but here each of the $w$ citations of each article is chosen preferentially for papers with higher citation counts, counted from the first to the current year. Therefore, this is similar to the Barabási-Albert model (see Refs.~\citep{ref14,ref15,ref17,ref16}), except by the fact that the in-degree (citation count) is not increased just after the addition of a new paper, but only at the end of a year. The DBPREF model is preferential both in terms of visibility and in-degree. More specifically, a list is kept where the identification of each article is entered a total number of times corresponding to the value of its citation count multiplied by the community visibility (we assume $f_A = 2$ and $f_B = 1$ in order to establish the proportion $f_A / f_B = 2$). New citations are then chosen by random, uniform selection among the elements in the above list.

\begin{figure}[h]
    \begin{center}
    \includegraphics[width=0.8\textwidth]{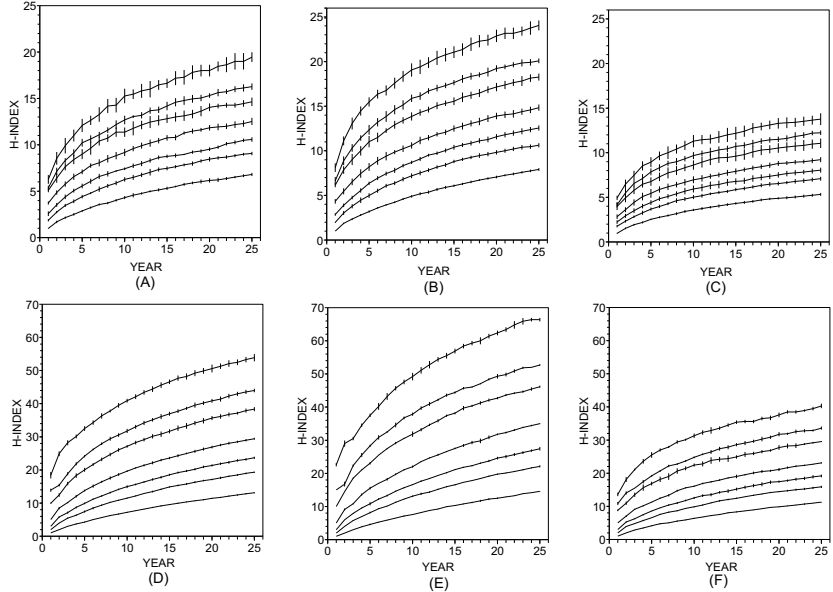}
    \end{center}
    \caption{\label{figres1} Dynamics of the h-index using (a) UNI model with $w=5$; (b) PREF model for community A with $w=5$; (c) PREF model for community B  with $w=5$; (d) UNI model  with $w=20$; (e) PREF model for community A with $w=20$; and (e) PREF model for community B with $w=20$.}
    \end{figure}

   Each of the configurations was performed $20$ times to provide statistical representativeness, while the h-index and total citation counts were computed for each author each year. The results in Figure \ref{figres1} indicate that including a preferential attachment (PREF) based on the fitness of a community has little effect for Community A, whose h-index increases marginally, but a large effect for Community B. Indeed, the h-index of all authors in Community B increased at a lower rate and after 25 years was considerably lower than that for authors in Community A. In fact, the h-index values are much smaller than for the model with random selection (UNI), as will be explained later on. These observations apply for $w$ = 5 or 20, though obviously the overall h-index values are higher for the networks built with $w$ = 20.

\begin{figure}[h]
    \begin{center}
    \includegraphics[width=0.8\textwidth]{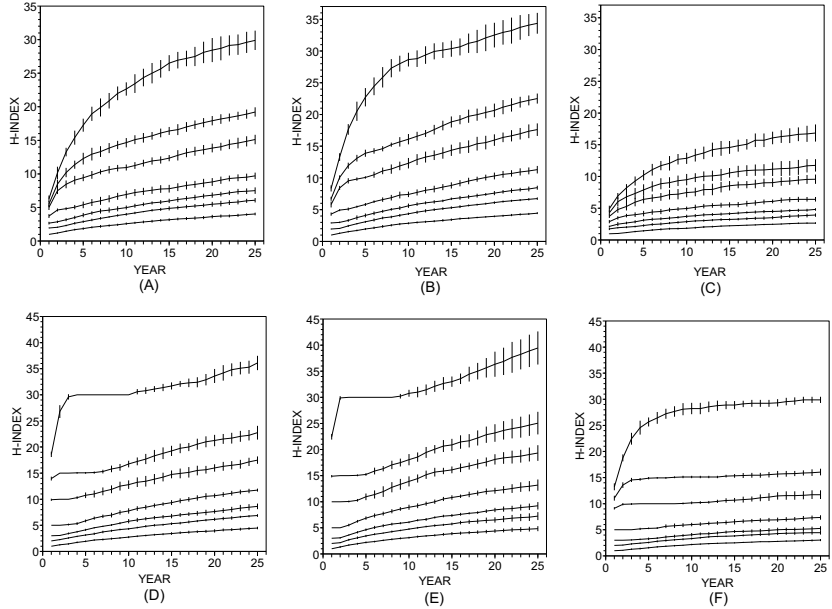}
    \end{center}
    \caption{\label{figres2} Dynamics of the h-index using (a) PREFC model with $w=5$; (b) DBPREF model for community A with $w=5$; (c) DBPREF model for community B with $w=5$; (d) PREFC model with $w=20$; (e) DBPREF model for community A with $w=20$; and (e) DBPREF model for community B with $w=20$.}
    \end{figure}

    With regard to the importance of citation counts, Figure \ref{figres2} shows a small increase in h-index for w = 5 in comparison with the UNI model (\ref{figres1}a). In contrast, the h-index values are much lower when applying the preferential attachment rule for w = 20 in Figure \ref{figres2}d than for the random case (UNI model in Figure \ref{figres1}d). When in addition to considering the in-degree (PREFC) we also consider the fitness (DBPREF), there is a marginal increase in h-index for community A, but the effects are again strong for community B. For the authors of the latter community, the h-index values achieved are much lower. The only exception appears to be for the author with the largest number of papers and w = 20. For some reason, there is a compensation effect in this case, and the h-index of this author is not so much lower. Note also that upon applying the preferential attachment rule based on the in-degree (for PREFC and DBPREF), the asymmetric distribution of citations among papers caused the h-index to be considerably lower than with the UNI or PREF models for a fixed number of references.

    \section{Conclusion} \label{sec7}

    The combination of the three factors, namely content similarity, in-degree and date of publication, has been effective in generating a model that reproduced several topological features of citation networks. The model represents, therefore, considerable progress compared to the literature in explaining the dynamics of citation networks. This applied to two real networks obtained from the arXiv repository for the topics ``graphenes'' and ``complex networks'', but the relative importance of the three factors varied for the networks. While the content similarity was the most relevant factor for both networks, the other two had distinct levels of importance depending on the network. {This network dependence probably highlights the expectation that other factors are also relevant for the pattern of citations as even highly similar articles can be forgotten~\cite{goodpractices}}. In fact, the reputation of authors and institutions is expected to play an important role, but its quantification is not possible with the current databases available. One may argue that the effect from reputation is at least partially taken into account when the in-degree is considered, for papers with larger citation count are more likely to receive additional citations~\citep{newmanbook,price}. But this is only an indirect manifestation of the reputation, which does not cover the higher visibility that papers from renowned authors and institutions have right after being published (when the citation count is still small or zero).  	 

    Owing to the importance of the visibility (or reputation) factor, we decided to verify its effects on the evolution of h-index of authors by considering artificial citation networks. For the latter we showed that the {community} with higher fitness (i.e. higher probability of having their paper being cited) - benefit only marginally - in terms of their h-index - in comparison with a control citation network with no bias. {This increase in the h-index of prominent authors probably occurs because the h-index is Lotkaian (follows a power law distribution) and therefore the concentration effect might be a reinforcement effect in three dimensional informetrics~\cite{eggs}.}  {In contrast, communities with less visibility can be hit hard, as their h-index values could be considerably lower than those estimated for the control, unbiased network. This finding confirms the observation in real networks that h-values depend on the productivity and citation practices of given fields~\cite{hin}. Therefore, caution should be taken when using the h-index to assess authors from distinct communities.}



    \section*{Appendix A - Setting Up the Parameters of the Model}

    Given the $3$ probability distributions concerning topological, semantic and temporal features, the model selects by chance one of these distributions to choose a paper to be included in its reference list. The prominence of each model is set according to the value of $2$ thresholds: $t_1 = \alpha$ and $t_2 = \alpha + \beta$. In other words, if the random number $n_r$, such that $0 \leq n_r \leq 1$, is less than $t_1$, then the $p(k)$ distribution is chosen. On the other hand, if $t_1 <n_r \leq t_2$, then the content similarity $p(\sigma)$ is chosen. Otherwise, if $t_2 <n_r \leq 1$, then $p(\Delta t)$ is selected. Thus, the prominence of the topological, semantic and temporal factors are given respectively by $\alpha$, $\beta$ and $\lambda=1-\alpha-\beta$.      In order to optimize the model, we minimized the following error $\epsilon^2$:
    \begin{equation}
    \begin{array}{lcl}
    \epsilon^2 & = & \frac{1}{6} \Bigg{(} \frac{\gamma_{k,m} - \gamma_{k,r}}{\gamma_{k,r}} \Bigg{)} ^ 2 + \frac{1}{12} \Bigg{(} \frac{\mu_{l,m} - \mu_{l,r}}{\mu_{l,r}} \Bigg{)} ^ 2 + \frac{1}{12} \Bigg{(} \frac{s_{l,m} - s_{l,r}}{s_{l,r}}  \Bigg{)}^2 +    \\ && \frac{1}{6} \Bigg{(} \frac{\mu_{\sigma,m} - \mu_{\sigma,r}}{\mu_{\sigma,r}} \Bigg{)} ^ 2 + \frac{1}{6} \Bigg{(} \frac{s_{\sigma,m} - s_{\sigma,r}}{s_{\sigma,r}}  \Bigg{)}^2 + \frac{1}{3} \Bigg{(} \frac{\gamma_{t,m} - \gamma_{t,r}}{\gamma_{t,r}} \Bigg{)} ^ 2,
    \end{array}
    \end{equation}
    whose parameters are explained in Table \ref{tab.1}.

    \begin{table}[h]
    \centering
    \caption{\label{tab.1} List of variables in the model.}
    \begin{tabular}{|c|c|}
    \hline
    \textbf{Variable} & \textbf{Meaning} \\
    \hline
    $\gamma_{k,m}$        &   Power law coefficient for the in-degree \\
    &   of the network obtained from the model. \\     $\gamma_{k,r}$        &   Power law coefficient for the in-degree in the real network. \\
    $\mu_{l,m}$           &   Average shortest path length of the network obtained from the model. \\
    $\mu_{l,r}$           &   Average shortest path length of the real network. \\
    $s_{l,m}$             &   Standard deviation of the shortest path \\
    &   length for the network obtained from the model. \\
    $s_{l,r}$             &   Standard deviation of the network obtained from the real network. \\
    $\mu_{\sigma,m}$      &   Average content similarity between an article and its references for the model.\\
    $\mu_{\sigma,r}$      &   Average content similarity between an article and its references for the real network.\\
    $s_{\sigma,m}$        &   Standard deviation of the content similarity \\
                          &   between an article and its references for the model.\\
    $s_{\sigma,r}$        &   Standard deviation of the content similarity between \\
                          &   an article and its references for the real network.\\
    $\gamma_{t,m}$        &   Power law coefficient of $\Delta_t$ for the network obtained from the model.\\
    $\gamma_{t,r}$        &   Power law coefficient of $\Delta_t$ for the real network.\\
    \hline
    \end{tabular}
    \end{table}

    The weights were distributed in order to give equal weighting to the three factors ($1/6 + 1/12 + 1/12 = 1/3$ for topology, $1/6 + 1/6 = 1/3$ for semantics and $1/3$ for the temporal feature). Because the brute-force search is impracticable, we made use of simulated annealing heuristic~\citep{recipes} in the simulations to minimize the error $\epsilon^2$.

    \section*{Acknowledgements}  	

    This work was supported by FAPESP and CNPq (Brazil).

\end{document}